\documentclass{ws-ijmpa}
\usepackage{ulem}
\usepackage{marginnote}  
\usepackage[super]{cite}
\usepackage{xcolor}
\usepackage[verbose,hypertexnames=false]{hyperref}
\hypersetup{colorlinks=false,allbordercolors=blue,pdfborderstyle={/S/U/W 1}}
\begin{document}

\markboth{Yao Chen, Ju Guo, Libo Zhao, Ziyao Yan, Jie Feng}{Advancing axion detection: Photon regeneration in high-sensitivity Penning trap experiments}

\catchline{}{}{}{}{}
%
%%%%%%%%%%%%%%%%%%%%%%%%%%%%%%%%%%%%%%%%%%%%%%%%%%%%%%%%%%%%%%%%%%%%

\title{Advancing axion detection: Photon regeneration in high-sensitivity Penning trap experiments}

\author{Yao Chen, Ju Guo, Libo Zhao}

\address{School of Instrument Science and Technology, Xi’an Jiaotong University, Xi’an 710049, China.
State Key Laboratory for Manufacturing Systems Engineering, International Joint Laboratory for Micro/Nano Manufacturing and Measurement Technologies, Xi'an Jiaotong University, Xi'an 710049, China.}

\author{Ziyao Yan, Jie Feng\footnote{
Corresponding author: fengj77@mail.sysu.edu.cn}}
\address{School of Science, Shenzhen Campus of Sun Yat-sen University, Shenzhen 518107, China}

\maketitle

\begin{history}
\received{18 July 2025}
\revised{16 June 2026}
\accepted{26 June 2026}
\published{}
\end{history}

\begin{abstract}

The axion, a hypothetical particle proposed to solve the strong CP problem and considered a viable candidate for dark matter, has prompted extensive experimental efforts for its detection. This study presents a novel approach combining photon regeneration techniques (``light-shining-through-a-wall") with high-sensitivity Penning trap technologies to enhance the search for axions. Penning traps offer significant advantages, including precise electromagnetic field measurement, strong magnetic fields, and single-particle detection capabilities. By integrating these traps with resonantly enhanced photon regeneration using microwave cavities, our proposed method significantly increases sensitivity to axion-photon couplings. Preliminary calculations demonstrate an unprecedented achievable sensitivity, reaching an axion-photon coupling constant limit of \(g_{a\gamma \gamma } \le 7.10\times 10^{-8}\mathrm{GeV ^{-1}} \) in just one day, specifically targeting axion energies below 1 MHz. This experimental setup presents a robust and controlled platform, circumventing astrophysical uncertainties, and represents a substantial advancement in laboratory searches for axions and our understanding of dark matter.

\end{abstract}


\section{INTRODUCTION}	

The observed gravitational velocity of a galaxy is greater than the expected one calculated from the visible masses, which is a hint of the existence of dark matter (DM) at the center of the galaxy. Another direct evidence of DM is its gravitational lensing effect together with X-ray observations. Apart from astronomical observations, cosmic microwave background (CMB) anisotropies are used to measure the portion of DM in the universe as 27\% \cite{Planck:2018vyg}.  

\begin{figure}
    \centering
    \includegraphics[width=\linewidth]{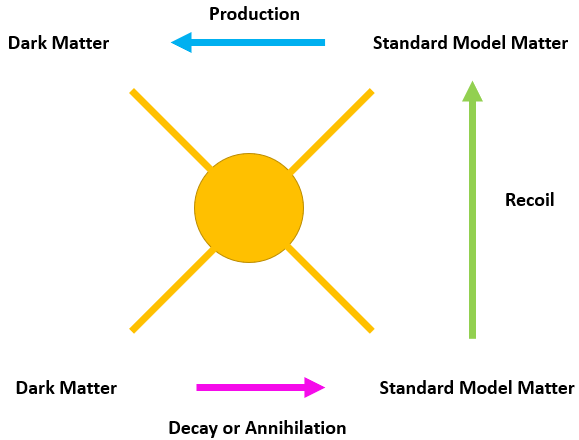}
    \caption{Dark matter detections are carried out in three complementary ways.}
    \label{fig:DM_detection}
\end{figure}

For astronomers, the dark matter (DM) problem is largely resolved through observational evidence. However, for particle physicists, the nature of the DM particle remains a profound mystery. Many models have been proposed. The mass range of the candidate is wide, we are studying the Axion-like particle. Dark matter is known to partake in gravitational interactions but intriguingly eludes electromagnetic interactions. The pursuit of understanding dark matter encompasses three complementary approaches, as illustrated in Fig.~\ref{fig:DM_detection}.

i) Studying the dark matter interactions, predominantly through indirect detection methods. These methods focus on identifying the by-products of DM particle annihilations or decays, such as gamma rays, neutrinos, or other standard model particles. Experiments like the Fermi Gamma-ray Space Telescope and the IceCube Neutrino Observatory exemplify this approach.

ii) Direct detection experiments, which aim to observe DM particles as they interact with ordinary matter. These experiments typically operate deep underground to shield from cosmic rays and rely on highly sensitive detectors waiting for the rare interactions between DM and the detector material. Experiments such as LUX, Xenon1T, and the future LZ experiment fall into this category.

iii) Collider experiments attempt to produce dark matter particles by colliding standard model particles at high energies. The Large Hadron Collider (LHC) at CERN is the most prominent example, where researchers analyze collision debris for signs of missing energy or momentum that might indicate the production of dark matter particles.

Within the framework of dark matter candidates, axionlike particles (ALPs) have garnered significant attention as a theoretically motivated subject of investigation. The Peccei-Quinn mechanism, a prominent theoretical construct, posits ALPs as a resolution to the strong CP problem in quantum chromodynamics (QCD) while simultaneously identifying them as potential constituents of dark matter. ALPs are postulated to exhibit ultralight masses and feeble couplings to Standard Model particles, most notably photons. This interaction forms the foundational basis for experimental detection methodologies, such as axion haloscopes (resonant cavity searches) and helioscopes (solar axion detectors).

In astrophysical contexts, constraints on ALP properties derived from solar axion emission inherently rely on solar models due to the influence of the Sun’s elevated plasma frequency and thermal environment, which perturb the effective axion-photon coupling.\cite{PhysRevLett.134.055001} Similarly, bounds inferred from stellar energy loss mechanisms—such as neutrino flux observations from Supernova 1987A—are subject to substantial astrophysical uncertainties.\cite{Bar_2020}

In contrast, laboratory-based experiments enable precise control over axion production processes, circumventing such model dependencies. Notable constraints have been established by accelerator-based studies, including the NOMAD (Neutrino Oscillation MAgnetic Detector)\cite{2000371}, Belle II\cite{Dolan_2017}, and NA64\cite{PhysRevLett.125.081801} experiments. While current laboratory searches do not yet achieve sensitivities comparable to astrophysical limits, they remain critically motivated by their immunity to astrophysical systematics. The pursuit of diverse experimental approaches remains imperative, as each methodology possesses distinct advantages and constraints in probing ALP parameter space.

This multifaceted strategy ensures robustness in exploring ALP physics, leveraging both astrophysical observations and controlled laboratory settings to advance our understanding of these elusive particles. \cite{PhysRevLett.134.055001}

Building upon these theoretical underpinnings and experimental endeavors, we introduces a novel approach: the use of photon regeneration techniques in Penning traps. In our study, axions can be detected through their ability to convert into radio-frequency (RF) photons in the presence of an external magnetic field. Therefore, the key to axion detection lies in achieving ultra-sensitive measurements of RF electromagnetic fields. We propose a novel axion detection scheme based on a Penning ion trap, which offers unique advantages for ultra-sensitive electromagnetic field sensing. In particular, studies show that the sensitivity for detecting electric fields around 1 MHz can exceed 1 $nV/m/Hz^{1/2}$ \cite{2021sciencepenning, phasecoherent2020}. Though talks about the use of Penning trap for axion detection have been developed, there is no such talk about the method of photon regeneration method in axion detection. Here, we focus on the usage of photon regeneration method for axion detection in Penning traps.

At this sensitivity level, the axion-photon coupling constant could be probed down to the order of $10^{-13} \mathrm{GeV^{-1}}$ or lower. By further increasing the number of ions and extending the measurement duration, the sensitivity to the coupling constant could be significantly enhanced, potentially surpassing the limits currently set by astrophysical observations. Compared with Rydberg atoms \cite{rydbergaxion2024}, the merit of using Penning traps includes: better sensitivity below 1 GHz, as well as a spare magnet is not needed due to the large confinement magnetic field in the Penning trap. Thus, the Penning-trapped ions are quite suitable for axion detection below 1 $\mu$eV. The sensitivity of axion detection based on Rydberg atoms could quickly decrease below the axion mass of 1 $\mu$eV.

The Penning ion trap can also be used for amplitude sensing\cite{2017amplitude}, quantum simulation\cite{2012britton}, and quantum information processing\cite{PhysRevXbilayercrystal2024,singlebitadressing}. In the area of large-scale ion arraying, a chip-based Penning ion trap system had been developed, achieving arbitrary ion arrangement\cite{jain2024penningmicronature}. Additionally, Penning traps operated at liquid helium temperatures can be used for the precision measurement of the antiproton magnetic moment and mass \cite{disciacca2013,2017smorra}, thus testing Charge conjugation- Parity transformation-Time reversal (CPT) symmetry violations. 

This paper is organized as follows. In Sect.~\ref{sec:methodologies}, we introduce the principle of the photon regeneration method to search for axions. Then we discuss the detail parameters of the experimental setup in Sect.~\ref{sec:setup}. Finally, we summarize our proposal in Sect.~\ref{sec:conclusion}.

\section{Methodologies}
\label{sec:methodologies}
\begin{figure}
    \centering
    \includegraphics[width=\linewidth]{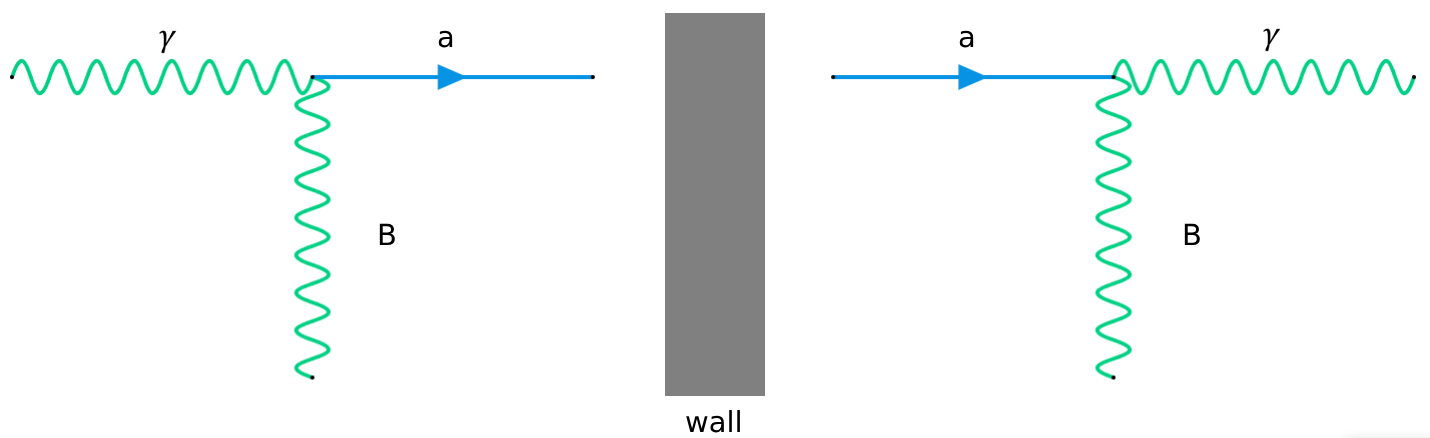}
    \caption{The schematic of the photon regeneration method.}
    \label{fig:Schematic}
\end{figure}
The Primakoff Effect (Primakoff effect) constitutes a pivotal mechanism in particle physics, describing the quantum process through which photons (photon) interacting with matter in intense electromagnetic fields can be converted into axions (axion) or axion-like particles (ALPs). This mechanism exhibits reversibility: under specific high-magnetic-field conditions, axions may undergo inverse conversion, regenerating detectable photons, as shown in Fig.~\ref{fig:Schematic}.

\begin{figure}
    \centering
    \includegraphics[width=\linewidth]{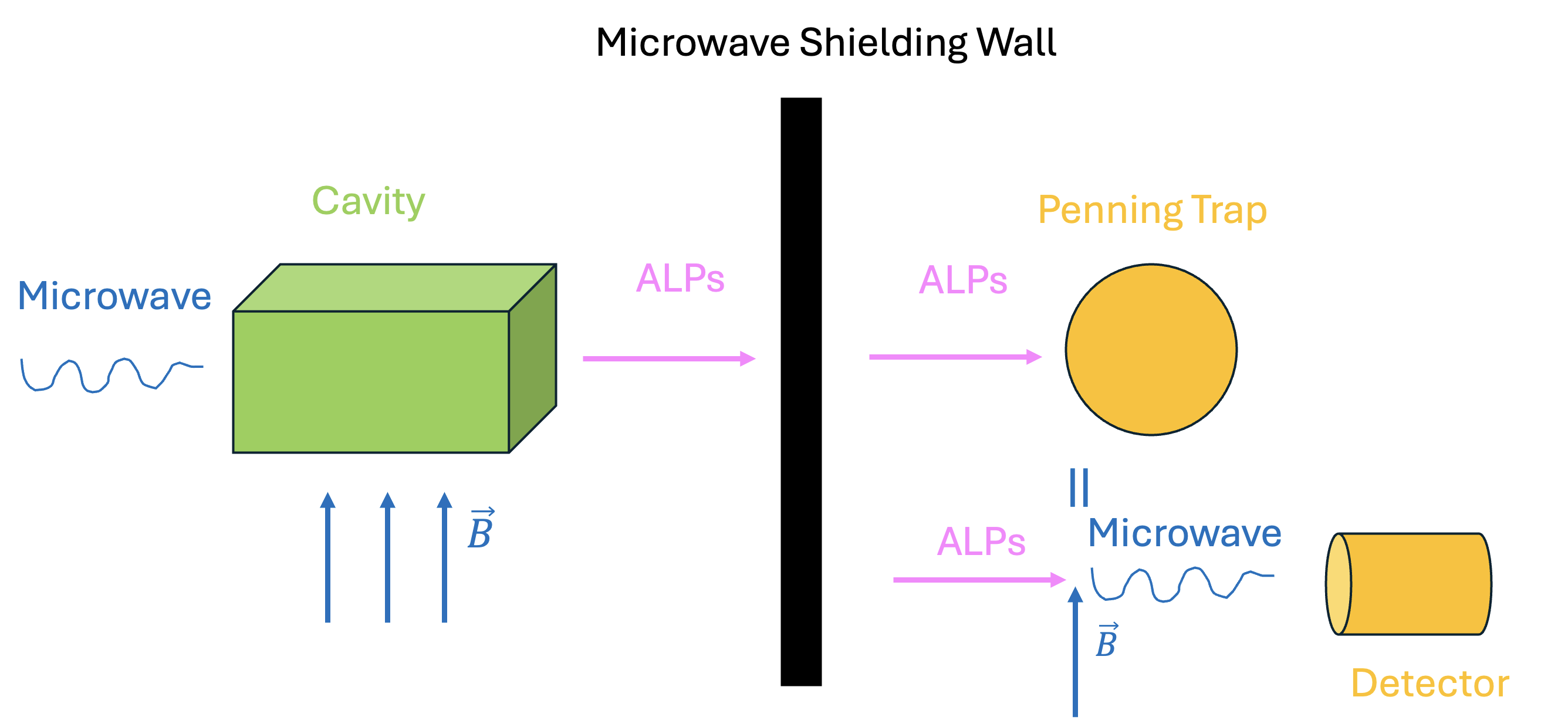}
    \caption{The experimental setup. A Microwave cavity to enhance the interaction between photons and the magnetic field is on the left. A Penning trap to increase the sensitivity of axion detection is on the right.}
    \label{fig:Setup}
\end{figure}

To enhance the interaction between the electric field of the photon and the external magnetic field, a microwave cavity is used at the generation site. 
On the other hand, to increase the sensitivity of the axion detection, we propose to use a Penning trap, which is a sensitive detector of the oscillating electromagnetic field that results from axion–photon conversion in the external static magnetic field, as shown in Fig.~\ref{fig:Setup}.
Externally injected photons are converted into axions via the Primakoff effect within the cavity. These axions subsequently traverse a photon-impenetrable wall and enter the detection chamber. This configuration ensures that all photons detected in the receiver cavity exclusively originate from axions undergoing inverse Primakoff effect regeneration.

\section{Experimental Setup}
\label{sec:setup}
\begin{figure}
    \centering
    \includegraphics[width=1\linewidth]{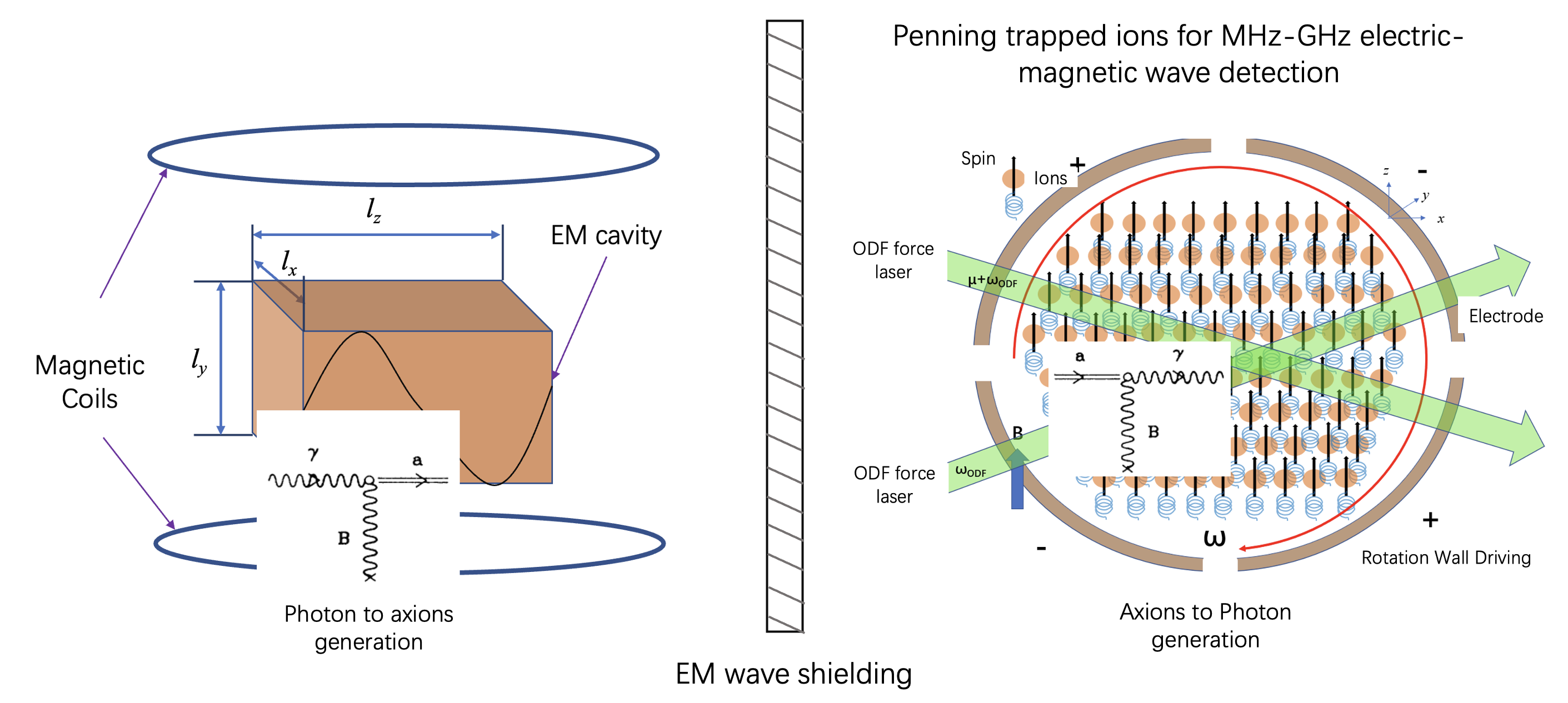}
    \caption{The experimental details of axion detection. ODF: Optical Dipole Force; EM: Electromagnetic.}
    \label{fig:priciple_experiment}
\end{figure}

The experimental details can be illustrated in Fig.~\ref{fig:priciple_experiment}, where the main instruments are shown. The parameters of the rectangular microwave cavity and the Penning trap are discussed separately in Sect.~\ref{subsec:microwave_cavity} and ~\ref{subsec:penning_trap}.

\subsection{Axion Production in the Microwave Cavity}
\label{subsec:microwave_cavity}

The rectangular microwave cavity in our experimental apparatus is precisely tuned to a resonant frequency of:
\begin{equation}
\omega _{\rm{mnp}} =\frac{\pi }{\sqrt{\mu \varepsilon}  }\sqrt{\left (  \frac{m}{l_{x}}\right )  ^{2}+  \left (  \frac{n}{l_y}\right )  ^{2}+  \left (  \frac{p}{l_z}\right )  ^{2} }.
\end{equation}

If \(l_z> l_x > l_y\), the lowest frequency mode is (1, 0, 1),
\begin{equation}
f _{\rm{101}} =\frac{1 }{2\sqrt{\mu \varepsilon}  }\sqrt{\left (  \frac{1}{l_x}\right )  ^{2}+  \left (  \frac{1}{l_z}\right )  ^{2} }.
\end{equation}

And its corresponding wavelength is

\begin{equation}
\lambda _{101} =\frac{2}{\sqrt{\left (  \frac{1}{l_x}\right )  ^{2}+  \left (  \frac{1}{l_z}\right )  ^{2} }}  .
\end{equation}

Its electric field component:

\begin{equation}
E_{\rm{x}} =0,
\end{equation}

\begin{equation}
E_{\rm{y}} =A_{2} \sin {\frac{\pi }{l_x} x}\sin {\frac{\pi }{l_z} z},
\end{equation}

\begin{equation}
E_{\rm{z}} =0.
\end{equation}

Under steady-state conditions, with a power of \( P \), the increase in energy per unit of time is \( P \). The quality factor of the microwave cavity is denoted by $Q$. The number of photons \( n \) generated per unit time satisfies the following:

\begin{equation}
QP=nhf_{101} .
\label{eq:n_photon}
\end{equation}

The efficiency of converting photon number to axion number is given by \cite{CAPPARELLI201637}:

\begin{equation}
P_{\gamma \to a} = g_{a\gamma\gamma}^2 B_{\rm{ext}}^2 \frac{\mathcal{E}_{\gamma}}{\sqrt{\mathcal{E}_\gamma^2 - m_a^2}} \left( \frac{\sin^2(q l/2)}{q^2}\right),
\end{equation}

where \( q = \mathcal{E}_{\gamma} - \sqrt{ \mathcal{E}_{\gamma}^{2} - m_{a}^{2} }\), \(B_{\rm{ext}} \) is the magnetic field strength, l is the interaction distance, \(\mathcal{E}_\gamma \equiv hf_{101}\) is photon energy, \(m_{a} \) is the mass of an axion, and \(g_{\alpha \gamma \gamma}\) is the dimensionless coupling constant between the axion and the photon.

We assume that the axion mass we are detecting is much smaller than the photon mass, we have \(\mathcal{E}_\gamma\gg m_{a}\) and \(q \approx \frac{m_{a}^{2}}{2 \mathcal{E}_{\gamma}}\). So the factor \(\frac{\mathcal{E}_\gamma}{\sqrt{\mathcal{E}_\gamma^2 - m_a^2}}\) $\simeq$ 1. The efficiency of converting photon number to axion number is \cite{PhysRevD.47.3707}:

\begin{equation}
P_{\gamma \to a} =\frac{l_z^{2}B_{\rm{ext}} ^{2} g_{a \gamma \gamma } ^{2}}{4},
\label{eq:a_efficiency}
\end{equation}

where $l_z$ is the effective interaction length. This expression is derived from the coherent conversion process of photons to axions in a magnetic field.

Taking Eqs.~\ref{eq:n_photon} and ~\ref{eq:a_efficiency} into account, the total mass of the axions produced per unit of time is given by:

\begin{equation}
M_{\rm{axion} } = n \times P_{\gamma \to a} \times m_a= \frac{l_z^2B_{\rm{ext}} ^{2} g_{a \gamma \gamma  } ^{2}QPm_{a} }{4hf_{101} } 
\end{equation}

The density of axions generated per unit time and per unit volume is given by:

\begin{equation}
\rho _{\rm{axion} } = \frac{M_{axion}}{l_xl_yl_z}= \frac{B_{\rm{ext}} ^{2} g_{a \gamma \gamma  } ^{2}QPm_{a}l_z }{4hf_{101} l_xl_y} .
\label{eq:a_density}
\end{equation}

\subsection{Principle of Electromagnetic Field Measurement Using a Penning Ion Trap}
\label{subsec:penning_trap}
As shown in Fig.~\ref{fig:priciple_experiment}, when multiple ions are confined using Penning trap technology under appropriately selected parameters, they form a two-dimensional planar ion crystal \cite{2021tang,2021paulcrystal,2015thompson}. A constant electric field is applied to the trap electrodes, and the entire trap is placed in a uniform magnetic field to achieve ion confinement\cite{gabrielse1986}. Unlike single-ion trapping, multi-ions could be confined. Each orange dot in the figure represents an individual ion, which can be envisioned as being attached to a tiny spring within the trap. 
The collection of these ions forms a planar structure, similar to a disk. The center-of-mass (COM) motion of this ion plane resembles the vertical oscillation (along the z-axis, parallel to the magnetic field direction) of a disk, with its vibrational frequency equal to the axial motional frequency\cite{2021sciencepenning}.

The entire ion crystal disk rotates at an angular velocity $\omega$, which can be precisely controlled through a rotating wall drive\cite{2017smorra}. By applying two ODF lasers, the spin states of the ions'electron can be entangled with their axial motional phonons. Using Ramsey-type spin direction detection methods, the COM mode amplitude of the ion vibrations can be measured via a quantum nondemolition (QND) measurement of the spin states\cite{2021sciencepenning}.

Because the ions carry electric charges, the presence of external electromagnetic fields—especially those with frequencies matching the ions’ axial motional frequencies (typically between 1 MHz and 1 GHz)—can induce resonant coupling. By detecting changes in the ions’ vibrational amplitude, one can sense the amplitude of external electromagnetic fields. To date, high-precision electric field measurements have been successfully demonstrated using this quantum mechanical oscillator\cite{2021sciencepenning}.
\subsection{Frequency tuning and ion cloud shape control}
In a Penning trap, the EM field detection frequency could be tuned through the trap voltage. Thus, the Penning-trapped ion system could also be called the axion radio. The frequency range is around 1 MHz-1 GHz for the typical design, which is quite different from other axion detection system. Thus, it is useful to talk about the frequency of the detection. Moreover, the shape of the ion cloud also could affect the detection. The key here is to control a ``disk" like shape. Thus, we also need to talk about the cloud shape.

A typical five-electrode cylindrical quadrupole configuration Penning trap design could be found in several references \cite{2020mcmahon,gabrielse1986}.The central ring electrode is divided into four sectors to implement a "rotating wall" drive, enabling control over the rotation rate $\omega$ of the ion ensemble. These segmented ring electrodes also serve as optical access ports for implementing laser cooling. The trapped ion ensemble naturally forms an ion crystal. This results from the interplay between the trapping forces of the ion trap and the Coulomb repulsion between ions, which causes the ions to arrange themselves in a lattice-like structure. Under typical parameter conditions, the inter-ion spacing is on the order of 10 micrometers, with an ion density around 10$^{15}/cm^3$.

As shown in the right part of Fig.~\ref{fig:priciple_experiment}, each ion is confined in the trap and can be equivalently modeled as being attached to a spring, characterized by an axial (z-direction) oscillation frequency $f_z$. The axial motion is decoupled from the radial and azimuthal modes, allowing the COM frame to be used to describe the collective axial vibrations of the ion crystal.

The overall shape of the ion crystal is determined by a combination of the trap voltages, external magnetic field strength, and the rotational angular velocity of the ion ensemble. In particular, by adjusting the rotating wall field, the rotation rate of the ions—and hence the shape of the crystal—can be tuned. For example, the crystal can be transformed from a prolate ellipsoid into a disk-shaped configuration.

Since the magnetic field of the Penning trap is generally fixed once the magnet is selected, the shape of the ion crystal must be manipulated by tuning the trap voltage and the rotation rate. Accordingly, it is essential to model the system in order to understand and predict the crystal shape under various operating conditions.

In the Penning trap, ions are confined using a static quadrupole electric potential of the form: $\Phi(r,t)=1/(4k_z)(2z^2-x^2 -y^2)$. $k_z$ is the effective axial spring constant determining the axial trapping frequency as $\omega_z=(qk_z/m)^{1/2}$, where q is the ion charge and m is the ion mass. Assuming a rotating wall potential is applied to stabilize and control the crystal rotation, the rotating electric potential is: $\Phi_{wall}(r,t)=1/(2k_z)\delta(x^2 -y^2)cos[2(\theta+ \omega_rt)]$, where $\delta$ is a dimensionless parameter representing the ratio between the rotating wall voltage and the static quadrupole voltage. To simplify analysis, it is convenient to move into the rotating frame of the wall potential, in which the ions appear stationary in steady-state. In this frame, the total potential energy of the system becomes:
\begin{equation*}
\Phi_r=\sum_{i=1}^{N}{\frac{1}{2}m\omega_z^2z_i^2}+\sum_{i=1}^{N}{\frac{1}{2}m\left[\omega_r\left(\omega_c-\omega_r\right)-\frac{1}{2}\omega_z^2\right]\left(x_i^2+y_i^2\right)}
+\sum_{i=1}^{N}{\frac{1}{2}m\delta\omega_z^2\left(x_i^2-y_i^2\right)}+\sum_{i=1}^{N}\sum_{j\neq i}^{N}{\frac{q^2}{8\pi\epsilon_0}\frac{1}{\mid\mathrm{r}_\mathrm{i}-\mathrm{r}_\mathrm{j}\mid}}.
\end{equation*}
where the subscript i refers to the i-th ion, $\omega_c=qB/m$ is the cyclotron frequency, and $\omega_r$ is the rotation frequency of the ion crystal. $x_i$,$y_i$, and $z_i$ are the position of the i-th ion.
To form a disk-shaped ion crystal, the radial confinement must be much weaker than the axial confinement. This can be quantified by introducing a dimensionless anisotropy parameter $\beta$:
\begin{equation}
\beta=\frac{\omega_r\left(\omega_c-\omega_r\right)-\left(1/2\right)\omega_z^2}{\omega_z^2}
\end{equation}
Achieving a planar, disk-shaped configuration requires $\beta \ll 1 $.
To achieve a precise modeling of the ion crystal shape, further considerations are necessary. Under typical conditions, the ion crystal assumes an ellipsoidal shape. One principal axis of the ellipsoid lies along the axial (z) direction, with a total length denoted as $2z_{cl}$, while the other principal axis lies in the z=0 plane, with a total diameter denoted as $2r_{cl}$. The aspect ratio $\alpha$ of the ellipsoid is then defined as: $\alpha=z_{cl}/r_{cl}$.
 
Through numerical simulations, the relationship between the shape of the ion ellipsoid and the trap parameters can be established. For instance, assuming the trapping magnetic field is 1 T, which could be easily achieved with the superconducting magnet. The Be ions are used due to the light atom weight of the ions. High axial frequency could be easily achieved with the tuning of the trap voltage. Suppose that the typical parameters of the 5-electrode are utilized, in which the diameter and the length of the electrodes are 5 mm and 5 mm, respectively. Under the trap voltage of 5000 V, the axial frequency could be around 10 MHz as well. We could easily change the rotation frequency of the ion cloud to achieve a disk-like shape of the ion crystal. Thus, the equivalence axion frequency around 10 MHz could be detected by our system as well, and the frequency could be tuned through the trap voltage. The trap voltage could be tuned to below 1 V, which means that the axion frequency below around 100 kHz could also be detected.
\subsection{Electromagnetic field induced amplitude measurement}
Since axions can convert into electromagnetic waves, they interact with charged ions and induce mechanical vibrations of the ion cloud in the Penning trap. Therefore, the detection of the amplitude of ion mechanical vibrations is a central objective. To achieve ultra-sensitive detection of ion vibrational amplitudes, a method based on spin-phonon entanglement could be used \cite{2017amplitude}. The ion vibrations can be modeled as quantum mechanical oscillators, with their excitations described by phonons.
To generate spin-phonon entanglement, an optical dipole force (ODF) is introduced. This is realized by two laser beams detuned from the ion’s internal transition frequency by several tens of GHz. The two lasers intersect in the ion region, creating a spatial gradient in the optical field. Due to the dipole interaction between the light field and the ion's electron, this gradient exerts a force on the ion. When the two lasers have a frequency difference $\mu$, an oscillating force is applied to the ions. Assuming the ion’s motion is given by $Z_ccos(\omega t+\delta_0)$, where $Z_c$ is the vibrational amplitude, $\omega$ is the vibration frequency, and $\delta_0$	is the vibration phase, the interaction Hamiltonian under the optical dipole force can be approximated as:
\begin{equation}
\widehat{H_{ODF}}=F_0Z_c\cos{\left(\Delta\mu t+\delta\right)}\sum_{i=1}^{N}\frac{\sigma_i^z}{2}
\end{equation}
Here, $\Delta\mu=\mu-\omega$ and $F
_0$ is the optical dipole force amplitude (tunable via laser power). $\sigma_i^z$ denotes the Pauli-Z operator for the 
i-th ion. When the ODF frequency matches the ion vibration frequency, the force induces an energy shift in the ion spin states, leading to Larmor precession. The precession angle is given by $\theta=\theta_{max}cos(\delta)$ where $\theta_{max}$ is equal to $\frac{F_0}{\hbar}Z_c\tau$, and $\tau$ is the interaction time. A Ramsey sequence can be used to measure $\theta$ \cite{phasecoherent2020}. Amplitude sensing could finally be changed into electromagnetic field sensing. Spin squeezing as well as increasing ion cloud numbers could further improve the sensitivity \cite{2021sciencepenning}. It is reported that electric-magnetic field sensing could reach below $1nV/m/Hz^{1/2}$ \cite{phasecoherent2020}.
\label{subsec:penning_trap}

In the simplified benchmark geometry, the generated axion field is assumed to consist of approximately equal downstream- and upstream-propagating components along the cavity axis. Only the downstream component is directed toward the shielding wall and the detection region. We therefore take the downstream geometric fraction to be 1/2 in the benchmark estimate. The electric field generated in the ion trap with a magnetic field $B_0$ is given by the following\cite{doi:10.1126/science.abi5226}:

\begin{equation}
E_{0} \sim (m_{\rm{a}} R)^2 g_{a \gamma \gamma } \alpha (t)B_{0},
\end{equation}

where:

\begin{equation}
\alpha \left ( t \right )  = \frac{\sqrt{\rho _{\rm{axion}}}}{m_{a} } \sin \left ( m_{a} t \right ) .
\end{equation}

Using a long-wavelength approximation, we are only interested in the magnitude of the electric field. The coupling constant sensitivity $ g_{a\gamma\gamma}$ is related to our electric field sensitivity $E_0$: 

\begin{equation}
E_{0} \sim (m_{\rm{a}} R)^2 g_{a \gamma \gamma }B_{0}\frac{\sqrt{ \frac{B_{\rm{ext}} ^{2} g_{a \gamma \gamma  } ^{2}QPm_{a}l_z }{4hf_{101} l_xl_y}}}{m_{a} }.
\label{eq:a_electricfield}
\end{equation}

The variation of both sides of Eq.~\ref{eq:a_electricfield} shows:

\begin{equation}
\delta g_{a\gamma \gamma } ^{2}  \sim  \left ( \delta E_{0} \right ) \frac{1}{\left ( m_{a}R \right ) ^{2}B_{0} B_{\rm{ext}} } \sqrt{\frac{4m_{a}hf _{101}l_xl_y}{PQl_z} }\sqrt{\frac{1}{t_{\rm{avg}}}}.
\end{equation}

In our experimental setup, we use the dimensions of the microwave cavity as 15.0 cm $\times$ 10.0 cm $\times$ 20.0 cm.
Using the current magnetic field, $B_{0} = 4.50$ T, $m_{a} = 4.14$ neV, assuming an electric field sensitivity $\delta E_{0}$ of \(10\rm{\frac{nV}{m}}\), the quality factor of the microwave cavity \(Q=1000\), \(P=1000\) W, $t_{\rm{avg}}=24\times3600$ s, $B_{\rm{ext}} = 1.00$ T and \(R=6.50\) cm (Here R denotes the characteristic transverse size of the magnetic-field region that contributes to the axion-induced electric field at the ion position, rather than the microscopic size of the ion crystal. The benchmark value $R=6.5\mathrm{cm}$ corresponds to the idealized magnet-bore geometry and should be understood as an optimistic reference case. The sensitivity scales as $\delta g_{a\gamma\gamma}\propto R^{-2}$.), we can reach the sensitivity at 1 day of averaging time:

%\begin{equation}
%JF\colon\delta g_{a\gamma \gamma }^{2}  \sim  \frac{6.52\times10^{-15}eV^2}{\left ( 4.14\times10^{-9} \times 3.29 \times 10^{5} \right ) ^{2} 878 eV^{2} \times 195 eV^{2} }\sqrt{ \frac{4\times4.14\times10^{-9}eV\times5.17\times10^{-6}eV\times3.80\times10^{5}eV^{-1}}{5.39\times10^{26}eV\times1000} } .
%\end{equation}

%\begin{equation}
  %JF\colon  \frac{6.52\times10^{-15}}{ (4.14\times10^{-9} \times 3.29 \times 10^{5})^{2}\times878 \times 195}\sqrt{\frac{4\times4.14\times10^{-9}\times5.17\times10^{-6}\times3.80\times10^{5}}{5.39\times10^{26}\times1000} }\rm{eV}^{-2}
%\end{equation}

\begin{equation}
\delta g_{a\gamma \gamma } ^{2}\left ( \rm{1day} \right ) %=5.04\times 10^{-33} \rm{eV^{-2}}
= 5.04\times 10^{-15} \rm{GeV^{-2}} ,
\end{equation}
where \(\hbar=c=1\). So 
\begin{equation}
\delta g_{a\gamma \gamma }\left ( \rm{1day} \right ) =7.10\times 10^{-8} \rm{GeV^{-1}} .
\end{equation}

\section{Conclusion}
\label{sec:conclusion}
In conclusion, this study introduces an innovative method for axion detection utilizing resonantly enhanced photon regeneration integrated with Penning trap technology. By leveraging the precise electromagnetic field measurement capabilities and the strong magnetic confinement provided by Penning traps, we propose an experiment capable of significantly advancing the sensitivity of axion searches in the energy range below 1 MHz. Our calculations demonstrate that a sensitivity level of axion-photon coupling as low as $\delta g_{a\gamma \gamma } =7.10\times 10^{-8} \rm{GeV^{-1}}$ can be achieved within just one day of operation, probing axion parameter spaces that have been previously inaccessible to laboratory experiments. This method circumvents the uncertainties inherent in astrophysical observations, offering a robust platform for controlled, precise exploration of axion physics. Continued refinement of this approach, including enhancements in ion number, measurement duration, and experimental design, holds substantial promise for uncovering the elusive axion and deepening our understanding of dark matter.

\section*{Acknowledgments}

This work is supported in part by National key research and development program under grant number 2022YFB3203400, "the Fundamental Research Funds for the Central Universities", the Guangdong Provincial Key Laboratory of Advanced Particle Detection Technology (2024B1212010005), the Guangdong Provincial Key Laboratory of Gamma-Gamma Collider and Its Comprehensive Applications (2024KSYS001), the Fundamental Research Funds for the Central Universities, and the Sun Yat-sen University Science Foundation.

\bibliography{sample}{}
\bibliographystyle{unsrt}

\end{document}